\documentclass[lettersize,journal]{IEEEtran}
\usepackage{graphicx}
\usepackage{caption}
\usepackage{subfigure}
\usepackage{amsmath}
\usepackage{amssymb}
\usepackage{bm}
\usepackage{color}
\usepackage{url}

\usepackage{graphicx,epstopdf}
\usepackage{epsfig}	
\usepackage{cite,graphicx,amsmath,amssymb}
\usepackage{comment}

\graphicspath{{./fig/}}

\begin{document}
\title{Integrated Sensing and Communications: A Mutual Information-Based Framework}
\author{Chongjun~Ouyang,~Yuanwei~Liu,~\IEEEmembership{Senior~Member,~IEEE,}\\ Hongwen~Yang,~\IEEEmembership{Member,~IEEE,}~and~Naofal~Al-Dhahir,~\IEEEmembership{Fellow,~IEEE}
\thanks{C. Ouyang and H. Yang are with the School of Information and Communication Engineering, Beijing University of Posts and Telecommunications, Beijing, 100876, China (e-mail: \{DragonAim,yanghong\}@bupt.edu.cn).}
\thanks{Y. Liu is with the School of Electronic Engineering and Computer Science, Queen Mary University of London, London, E1 4NS, U.K. (e-mail: yuanwei.liu@qmul.ac.uk).}
\thanks{N. Al-Dhahir is with the Department of Electrical and Computer Engineering, The University of Texas at Dallas, Richardson, TX 75080 USA (e-mail: aldhahir@utdallas.edu).}
}

\maketitle

\begin{abstract}
Integrated sensing and communications (ISAC) is potentially capable of circumventing the limitations of existing frequency-division sensing and communications (FDSAC) techniques. Hence, it has recently attracted significant attention. This article aims to propose a unified analytical framework for ISAC from a mutual information (MI) perspective. Based on the proposed framework, the sensing performance and the communication performance are evaluated by the sensing MI and the communication MI, respectively. The unity of this framework is originated from the fact that the sensing and communication (S\&C) performance metrics, i.e., the S\&C MI, have the similar physical and mathematical properties as well as the same unit of measurement. Based on this framework, the S\&C performance of downlink and uplink ISAC systems is investigated and compared with that of FDSAC systems. Along each considered system settings, numerical results are provided to demonstrate the superiority of ISAC over conventional FDSAC designs. Finally, promising open research directions are provided in the context of MI-based ISAC.
\end{abstract}

\section{Introduction}
\begin{figure*}[!t]
\centering
\setlength{\abovecaptionskip}{0pt}
\includegraphics[width=0.5\textwidth]{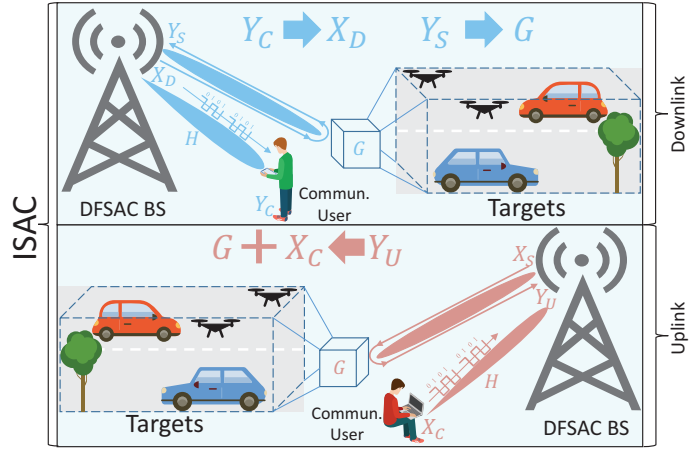}
\caption{Illustration of downlink and uplink ISAC systems. For downlink ISAC, $\emph{X}_{\emph D}$ is the DFSAC signal, $\emph{H}$ is the communication channel, $\emph{Y}_{\emph{C}}$ is the received signal at the communication user, $\emph{G}$ is the target response, and $\emph{Y}_{\emph S}$ is the reflected signal. For uplink ISAC, $\emph{X}_{\emph C}$ is the communication signal, $\emph{H}$ is the communication channel, $\emph{X}_{\emph{S}}$ is the sensing probing signal, $\emph{G}$ is the target response, and $\emph{Y}_{\emph U}$ is the received superposed signal at the BS.}
\label{System_Model}
\end{figure*}

Next-generation wireless networks are envisioned to play a big part in shaping a connected, smart, and intelligent wireless world \cite{Saad2019}. This will require a paradigm shift in future networks to support high-quality wireless connectivity as well as high-accuracy sensing capability \cite{Liu2022}. To fulfill this dual-functional requirement, numerous potential technologies have been introduced over the last few years. Among them is integrated sensing and communications (ISAC), a technique to achieve dual-functional sensing and communications (DFSAC) via a single time-frequency-power-hardware resource \cite{Liu2022,Zhang2022,Zhang2021,Liu2021}. Two fundamental ISAC models are illustrated in {\figurename} {\ref{System_Model}}, where the DFSAC base station (BS) aims to serve uplink/downlink communication users (CUs) while simultaneously sensing the targets in its surrounding environment. Compared to the frequency-division sensing and communications (FDSAC) techniques, in which sensing and communications (S\&C) require isolated frequency bands as well as hardware infrastructures, ISAC is capable of improving the spectral efficiency, reducing the hardware cost, and limiting the electromagnetic pollution \cite{Liu2022,Zhang2022,Zhang2021,Liu2021}. In view of the above benefits, ISAC has attracted vibrant industrial and academic interest, which is anticipated to dominate the future wireless network market.

Recent years have witnessed a growing number of published papers around the ISAC theme. These works can be roughly categorized into two main topics, including S\&C performance analysis \cite{Liu2021} and dual-functional beamforming design \cite{Liu2022_2,Yuan2021_1,Yuan2022_2}. Around these two topics, many S\&C performance metrics have been proposed. In most current ISAC research, the sensing performance is evaluated from an estimation-theoretic view, while the communication performance is evaluated from an information-theoretic view \cite{Liu2021,Liu2022_2,Yuan2021_1,Yuan2022_2}. Nevertheless, most estimation theory (ET)-based performance metrics involve different units of measurement, expressions, physical meanings, and mathematical properties from the information theory (IT)-based performance metrics. Therefore, current research does not provide a unified analytical framework for ISAC. Here, the so-called unified framework means that the S\&C performance metrics therein have the similar physical and mathematical properties as well as the same unit of measurement. We believe that a unified ISAC framework will help simplify the S\&C performance analysis and dual-functional beamforming design, which will be detailed in Section \ref{Section2B}.

Against the above background, we focus on the S\&C performance in ISAC systems. The new contributions of this article are that we establish a unified as well as analytically tractable framework for ISAC from a mutual information (MI) perspective. Compared to other existing ISAC frameworks, our proposed MI-based framework yields more theoretical generality and computational tractability. On this basis, we introduce several MI-related S\&C performance metrics and rely on two of them to study the performances of downlink and uplink ISAC systems. Numerical results are provided to demonstrate the superiority of ISAC over conventional FDSAC designs. Finally, conclusions are drawn and open research problems are highlighted.

\section{MI-Based Framework for ISAC}

\subsection{Mutual Information in ISAC}
\begin{figure*}[!t]
\centering
\setlength{\abovecaptionskip}{0pt}
\includegraphics[width=0.6\textwidth]{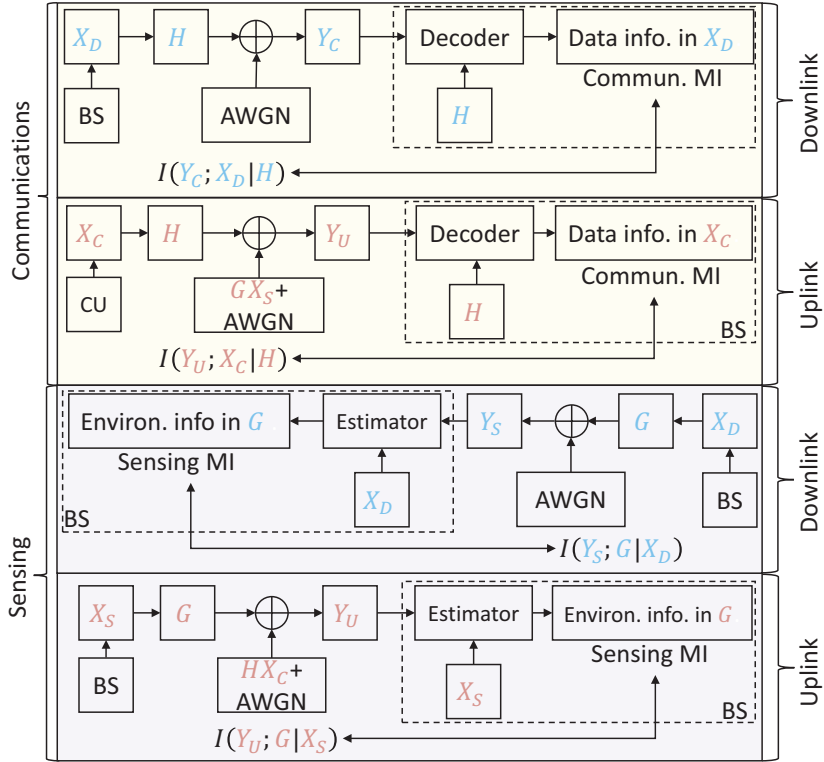}
\caption{Signal processing diagrams and S\&C MI in downlink and uplink ISAC systems. The notation $I(A;B|C)$ represents the mutual information between $A$ and $B$ conditioned on $C$.}
\label{S_C_MI}
\end{figure*}

As mentioned earlier, ISAC is envisioned to support dual-functional S\&C in the same frequency-time resource block. Generally speaking, communication aims at data information transmission, while sensing aims at environmental information extraction \cite{Liu2022}, as detailed in {\figurename} {\ref{S_C_MI}}. Despite having different direct objectives, both communications and sensing process \emph{information}. Since these two functionalities involve \emph{information} as their common core, one would naturally think of evaluating the S\&C performance from an information-theoretic view.

It is widely recognized that MI is the signature figure of merit in information theory. Particularly, the MI in ISAC can be categorized into communication MI and sensing MI.
\subsubsection{Communication MI} Essentially, the objective of communications is to recover data information contained in the transmitted symbol from the received signal. To this end, the communication channel is usually assumed to be known by the receiver, with which the data information can be successfully decoded \cite{Tse2005}. To clarify this point, in {\figurename} {\ref{S_C_MI}}, we illustrate the processing of communication signals in both uplink and downlink ISAC. As shown, in the communication part of ISAC, the receiver leverages the communication channel ($\emph{H}$) to recover data information contained in the transmitted symbols (${\emph{X}}_{\emph C}$ or ${\emph{X}}_{\emph D}$) from the received signal (${\emph{Y}}_{\emph U}$ or ${\emph{Y}}_{\emph C}$). On this basis, we define the communication MI as the MI between the transmitted symbol and the received signal conditioned on the communication channel. This metric evaluates the information-theoretic limit on how much data information can be transmitted without error \cite{Tse2005}. As is widely known, for a communication channel with Gaussian signalling, the communication MI can be calculated by the Shannon formula and be maximized by water-filling solutions \cite{Xing2020}.
\subsubsection{Sensing MI}
Let us turn now to sensing whose objective is to reconstruct the surrounding environment. To this end, a predesigned sensing probing signal should be firstly broadcasted to the nearby environment and then the environmental information can be recovered from reflected echoes. Typical environmental information includes the number, directions, distances, and speeds of the targets, which together constitute a target response $\emph{G}$ \cite{Tang2019}, as depicted in {\figurename} {\ref{System_Model}}. Particularly, with $\emph{G}$ at hand, one can reconstruct the surrounding environment in real time, since $\emph{G}$ contains all the necessary environmental information. Essentially, the objective of sensing is to estimate the target response $\emph{G}$. Taken together, from an information-theoretic view, sensing aims at extracting environmental information contained in the target response from reflected echoes with a predesigned sensing probing signal. To better show this, in {\figurename} {\ref{S_C_MI}}, we illustrate the processing of sensing signals for both uplink and downlink ISAC systems. As shown, in the sensing part of ISAC, the BS leverages the predesigned sensing probing signal ($\emph{X}_{\emph{S}}$ or $\emph{X}_{\emph{D}}$) to recover environmental information contained in the target response ($\emph{G}$) from the reflected signal (${\emph{Y}}_{\emph U}$ or ${\emph{Y}}_{\emph S}$). To quantify how much environmental information can be extracted with a given probing signal, we define sensing MI as the MI between the target response and the reflected signal conditioned on the predesigned sensing probing signal, as depicted in {\figurename} {\ref{S_C_MI}}. It is worth noting that the sensing procedure is generally characterized by a linear Gaussian model under which the sensing MI can be also calculated by the Shannon formula \cite{Tang2019}. Furthermore, as discussed in \cite{Ouyang2022_2}, the sensing MI can be also maximized via the water-filling solution.
\subsubsection{MI-Based ISAC Framework}
The above discussions suggest that the S\&C MI satisfies the following properties:
\begin{itemize}
  \item The sensing MI has the similar physical meaning to the communication MI, both characterizing the information-theoretic limits on how much information can be recovered.
  \item The sensing MI has the same unit of measurement as the communication MI. For example, both of them are measured by bits when using the binary logarithm.
  \item The sensing MI has a similar mathematical expression to the communication MI, both being calculated by the Shannon formula.
  \item The sensing MI has similar optimization methods to the communication MI, both being improved by traditional optimization methods for MI, such as the water-filling method and its numerous variants.
\end{itemize}
By using S\&C MI as the S\&C performance metrics, we establish a MI-based ISAC framework. Under this framework, the S\&C performance metrics have the similar physical and mathematical properties as well as the same unit of measurement. This is in contrast to most current ISAC research where the sensing performance and the communication performance are evaluated by ET-based and IT-based metrics, respectively. Therefore, it can be concluded that this MI-based ISAC framework is more unified than other existing frameworks. This unified framework allows a more robust and simpler evaluation of the S\&C performance.

\subsection{Estimation-Theoretic View vs. Information-Theoretic View}\label{Section2B}

In the above, we established the basic concepts of S\&C MI as well as the MI-based ISAC framework. To elaborate further, in the following subsection we explain in detail our motivation of establishing this MI-based framework.

Generally, the S\&C performance in ISAC can be evaluated from either an estimation-theoretic view or an information-theoretic view. From the perspective of estimation, the sensing performance is evaluated using the Cram\'{e}r-Rao bound (CRB), target detection probability (TDP), beampattern matching error (BME), and mean squared error (MSE), while the communication performance is evaluated using the bit error rate (BER), symbol error rate (SER), and MSE \cite{Liu2022}. On the other hand, from an information-theoretic view, the sensing performance and the communication performance can be characterized by the sensing MI and the communication MI, respectively.

Upon comparing the ET-based metrics with the IT-based metrics, we note that the ET-based metrics are generally more intuitive than the IT-based ones. This is because the ET-based metrics explicitly quantify the recovery accuracy of both data information and environmental information. However, it should be pointed out that the expressions of most ET-based metrics, such as the CRB and MSE, have complicated forms, which are more computationally and analytically involved than the IT-based ones \cite{Tang2019}. Moreover, many ET-based metrics are based on specific estimation algorithms. For example, the CRB relies on an unbiased maximum-likelihood estimator. For another example, the TDP relies on a statistical hypothesis tester. This is in contrast with the IT-based metrics, which do not rely on specific estimation methods and thus are more general. Last but not least, some ET-based metrics, such as the BME and TDP, have a weak connection with the MI. Therefore, these ET-based metrics cannot be used to improve the MI of ISAC systems. In contrast to this, optimizing the IT-based metric can also improve the estimation performance. Particularly, as suggested in \cite{Tang2019}, maximizing the sensing MI can also: 1) minimize the MSE in estimating the target response matrix, 2) achieve the Chernoff bound in sensing uncorrelated targets, and 3) improve the detection probability. Based on the comparison and analysis above, we conclude that the IT-based metrics yield more computational tractability and theoretical generality than the ET-based ones. Therefore, using the IT-based metrics, i.e., S\&C MI, as the S\&C performance metrics not only makes the subsequent dual-functional beamforming design more tractable but also makes the derived analytical results more general.

As stated before, in conventional ISAC research, the sensing performance and the communication performance are often evaluated by the ET-based and the IT-based metrics, respectively. In general, the ET-based metrics have different units, expressions, physical meanings, and mathematical properties from the IT-based metrics. This could make the resulting S\&C performance analysis frustratingly difficult. Actually, this is also the reason why the performance gap between ISAC and FDSAC has not been fully understood. Besides, as stated before, the ET-based metrics are computationally intractable and lack sufficient theoretical generality, which could complicate the corresponding beamforming design. In view of the above shortcomings, we propose to use the IT-based metrics, i.e., S\&C MI, to evaluate the S\&C performance, which have the similar physical and mathematical properties as well as the same unit of measurement. Besides, the S\&C MI is more computationally tractable and theoretically general than the ET-based metrics. Consequently, by resorting to the S\&C MI, we succeed in establishing a unified as well as analytically tractable framework for ISAC.

Based on the above arguments, it can be concluded that the proposed MI-based framework yields more theoretical generality and computational tractability than other existing ISAC frameworks. In view of these benefits, this MI-based framework is envisioned to facilitate the S\&C performance analysis and the dual-functional beamforming design.

\subsection{Proposed MI-Related Performance Metrics for ISAC}
By virtue of the proposed MI-based framework, we can analyze the fundamental S\&C performance to unveil important system design insights. Also, this framework enables us to compare the performances between ISAC and FDSAC techniques. In the sequel, we introduce several commonly used MI-related S\&C performance metrics.
\begin{itemize}
  \item \textbf{Sensing-Communication (S-C) Rate Region}: The communication rate (CR)/sensing rate (SR) is defined as the communication MI/sensing MI achieved in a unit time-frequency resource block. The S-C rate region defines a set containing all the achievable SR-CR pairs, which explicitly characterizes how much information can be transmitted or extracted.
  \item \textbf{High-Singla-to-Noise Ratio (SNR) Slopes}: The high-SNR slope is also known as the degree-of-freedom, the rate pre-log, or the multiplexing gain \cite{Tse2005}. The high-SNR slope of the CR/SR is calculated by taking the high-SNR limitation of the ratio of the CR/SR to the logarithm of the SNR. Thanks to its analytical tractability, this notion is an effective tool to evaluate the S\&C performance in the high-SNR regime.
\item \textbf{Other MI-Related Performance Metrics}: In addition to the S-C rate region and high-SNR slope, other MI-related metrics, such as the high-SNR power offset, outage probability, diversity order, and rate distortion function, can be used to glean further system insights.
\end{itemize}

\section{Mutual Information in Downlink ISAC}
The previous section laid a solid foundation for understanding the MI in ISAC systems. It is now necessary to discuss ISAC systems from a MI perspective. To be more specific, our aim is to evaluate the S\&C performance of ISAC with two MI-related metrics, i.e., the S-C rate region and high-SNR slope. Our hope is that this performance evaluation will contribute to a deeper understanding of the superiority of ISAC over conventional FDSAC. In this section, we focus on the MI in downlink ISAC systems, as depicted in the right side of {\figurename} {\ref{System_Model}}. Downlink ISAC refers to an ISAC scheme where the BS transmits data to downlink CUs while simultaneously sensing the targets in its surroundings.

\subsection{Rate Region Characterization}
\begin{figure}[!t]
\centering
\setlength{\abovecaptionskip}{0pt}
\includegraphics[width=0.45\textwidth]{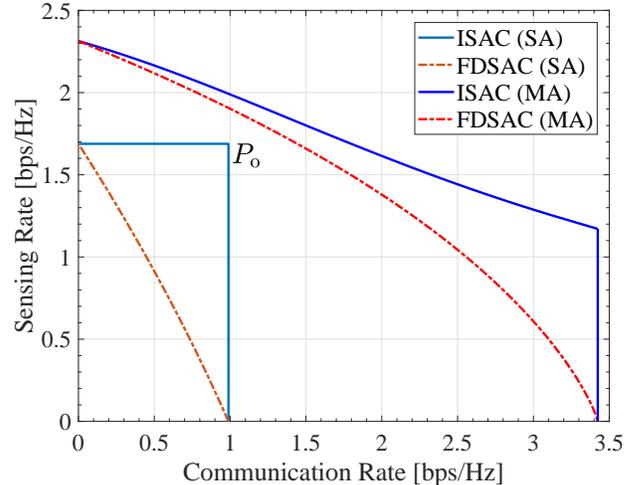}
\caption{Communication-sensing rate region achieved in downlink ISAC and FDSAC. The system parameters and other pertinent details can be found in \cite{Ouyang2022_2} and \cite{Ouyang2022_1}.}
\label{Downlink_Region}
\vspace{-10pt}
\end{figure}

Let us first study the S-C rate region of the downlink ISAC system. Particularly, the rate region is characterized in two steps. In the first step, we assume that the BS uses a single transmit antenna to serve a pair of single-antenna (SA) CUs as well as sensing the targets \cite{Ouyang2022_1}. Under this scenario, there is no need to consider transmit beamforming, and it is desirable to transmit at the maximum power to maximize either the CR or the SR \cite{Ouyang2022_1}. Additionally, nonorthogonal multiple access (NOMA) is exploited to serve the CUs because of inter-user interference mitigation capability. It is noteworthy that from a communication perspective, NOMA can also achieve the capacity region of this two-CU downlink channel \cite{Liu2017}. The above system settings enable us to generate a DFSAC signal that maximizes the CR and SR at the same time \cite{Ouyang2022_1}. Using this signalling, both communications and sensing can enjoy all the power-spectrum resources and there is no inter-function interference as well as S\&C performance trade-offs. This is in contrast to the FDSAC, in which both communications and sensing only benefit from partial power-spectrum resources. It is thus foreseen that ISAC is capable of achieving a broader C-S rate region than FDSAC. To illustrate this point, we compare the rate regions achieved by ISAC and FDSAC under the SA case in {\figurename} {\ref{Downlink_Region}}. What stands out in this graph is that the rate region achieved by FDSAC is completely included in that achieved by ISAC. At the point $P_{\rm{o}}$, ISAC can attain both the maximum CR and the maximum SR; therefore, this point serves as supremum of the whole system.

The performance of ISAC can be further boosted in multi-antenna networks. Thus, in the second step, we discuss a multiple-antenna (MA) case \cite{Ouyang2022_2}. In this case, transmit beamforming design has a significant impact on the S\&C performance. On the one hand, the SR-optimal beamformer can be calculated by using the water-filling method \cite{Ouyang2022_2}. On the other hand, the sum-CR-optimal beamformer can be calculated via capitalizing on the dirty-paper coding (DPC) and uplink-downlink duality as well as the iterative water-filling method \cite{Ouyang2022_2}. Therefore, finding a beamforming strategy to maximize the CR and the SR at the same time is an arduous task. In other words, it is challenging to make communications and sensing simultaneously take full use of all the power resources. This would result in mutual power leakage between these two functionalities. Due to this leakage, the MA case suffers from inter-function interference.

For analytical tractability, we propose a simple superposition-based beamforming design \cite{Ouyang2022_2} to mitigate the inter-function interference. Specifically, we first generate the sensing signal by taking account of the statistical properties of the communication channel. Then, we exploit the DPC to generate the communication signal in order to maximize the sum CR. Finally, we generate the DFSAC signal by superposing the designed communication and sensing signals. Afterwards, the BS broadcasts the DFSAC signal to the CUs and STs. Since the sensing signal requires no instantaneous information of the communication channel, it can be designed in advance and fed back to the CUs. After the CUs receive the DFSAC signal, they will firstly remove the sensing signal therein and then decode the date information contained in the communication signal. Meanwhile, the BS receives the echoes reflected from the targets. By treating the communication signal contained in the echo signal as interference, the BS will extract the environmental information contained in the target response matrix. Compared to FDSAC, the superposition-based ISAC enables both communications and sensing to make full use of the entire spectrum resources. Hence, the superposition-based ISAC is capable of achieving a broader S-C rate region than FDSAC \cite{Ouyang2022_2}. To better show this benefit, in {\figurename} {\ref{Downlink_Region}}, we compare the rate regions achieved by FDSAC and the superposition-based ISAC under the MA case. As expected, it can be seen from {\figurename} {\ref{Downlink_Region}} that ISAC yields a broader rate region than FDSAC. Moreover, by comparing the rate regions achieved by the SA and MA cases, one can conclude that a proper beamforming design can effectively broaden the rate region for both FDSAC and ISAC. Last but not least, as {\figurename} {\ref{Downlink_Region}} shows, under the MA case, there is no point that can simultaneously attain the maximum CR and SR. This is also the reason why the rate region achieved by ISAC under the MA case is not a rectangular one.

\subsection{High-SNR Slopes}

\begin{figure}[!t]
\centering
\setlength{\abovecaptionskip}{0pt}
\includegraphics[width=0.45\textwidth]{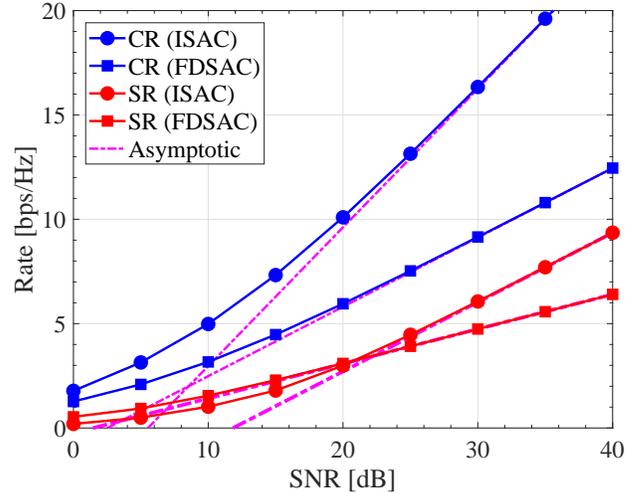}
\caption{Communication and sensing rates for downlink ISAC and FDSAC. The system parameters can be found in \cite{Ouyang2022_2}.}
\label{Downlink_Rate}
\vspace{-10pt}
\end{figure}

Having characterized the C-S rate region, we now move to the high-SNR slopes. The high-SNR slope represents the slope with which the CR/SR curve increases with the SNR in the high SNR regime. A larger high-SNR slope will lead to a better CR/SR performance in the high-SNR region. This subsection aims to unveil the performance gap between ISAC and FDSAC by using the high-SNR slope as a metric.

Particularly, isolated frequency bands are used for sensing and communications in FDSAC; therefore, a spectrum allocation factor smaller than one will appear in the pre-log term of the corresponding CR/SR expression. By contrast, all the spectrum resources in ISAC can be used for communications as well as sensing. Thus, the pre-log term of the corresponding CR/SR expression will not be decreased by a factor less than one. The above argument implies that ISAC achieves larger high-SNR slopes than FDSAC in terms of both CR and SR. To highlight this point, we first study the SA case where the BS is equipped with a single transmit antenna. As stated before, in this case, a DFSAC signal is generated to concurrently attain the maximum CR and SR. Thus, making ISAC enjoy a higher CR as well as SR than FDSAC is always possible, even for a sufficiently large SNR. This, essentially, shows that ISAC yields larger high-SNR slopes than FDSAC \cite{Ouyang2022_1}. Turning now to the MA case where the BS is equipped with multiple transmit antennas for beamforming. As mentioned previously, all the spectrum resources are shared by communications and sensing together in the proposed superposition-based ISAC. Hence, the superposition-based ISAC is anticipated to bring in larger high-SNR slopes than FDSAC. To quantify this benefit, in {\figurename} {\ref{Downlink_Rate}}, we plot the CR and SR with respect to the SNR under the MA case. Notice that the high-SNR slope of CR/SR is determined by the slope of the asymptotic CR/SR curve. Hence, it can be concluded from {\figurename} {\ref{Downlink_Rate}} that ISAC brings in larger high-SNR slopes than FDSAC in terms of both CR and SR. As {\figurename} {\ref{Downlink_Rate}} shows, ISAC achieves a higher CR than ISAC for all the considered SNR ranges. By contrast, in the low and moderate SNR regions, FDSAC is capable of achieving a higher SR than ISAC. This is because the communication signal interferes in ISAC’s sensing procedure, thus reducing the SR. Yet, due to its larger high-SNR slope, the SR of ISAC will exceed that of FDSAC as the SNR increases, which agrees with the observations from {\figurename} {\ref{Downlink_Rate}}.

\section{Mutual Information in Uplink ISAC}
Another fundamental ISAC model is termed as the uplink ISAC, where the DFSAC BS aims to extract environmental information from the reflected sensing echoes while simultaneously detecting the data symbols sent by uplink CUs, as illustrated in the left side of {\figurename} {\ref{System_Model}}. By resorting to the MI as a metric, the performance gap between ISAC and FDSAC can be quantified. 

\subsection{Rate Region Characterization}
To begin with, we characterize the rate region achieved in the uplink ISAC. In an effort to deal with the superposed sensing and communication signals, we establish a successive interference cancellation (SIC)-based framework \cite{Ouyang2022_2,Ouyang2022_3}. Specifically, two SIC schemes are proposed, i.e., the sensing-centric (S-C) SIC and the communications-centric (C-C) SIC, which correspond to two different SIC orders. In the S-C SIC scheme, the BS first decodes the communication signal by treating the sensing signal as interference. Then, the communication signal is subtracted from the superposed signal and the remaining part will be used to sense the target response. As for the C-C SIC, the BS will first estimate the target response matrix by treating the communication signal as interference. Afterwards, the sensing signal is subtracted from the superposed signal and the remaining part will be used to recover the data information sent by the CUs. For simplicity, an ideal SIC condition is considered here, where the interference is assumed to be perfectly cancelled. It is clear that in the S-C SIC scheme, the communication signal has no influence on the sensing procedure due to interference cancellation, whereas in the C-C scheme, the sensing signal has no influence on the communication procedure.

\begin{figure}[!t]
\centering
\setlength{\abovecaptionskip}{0pt}
\includegraphics[width=0.45\textwidth]{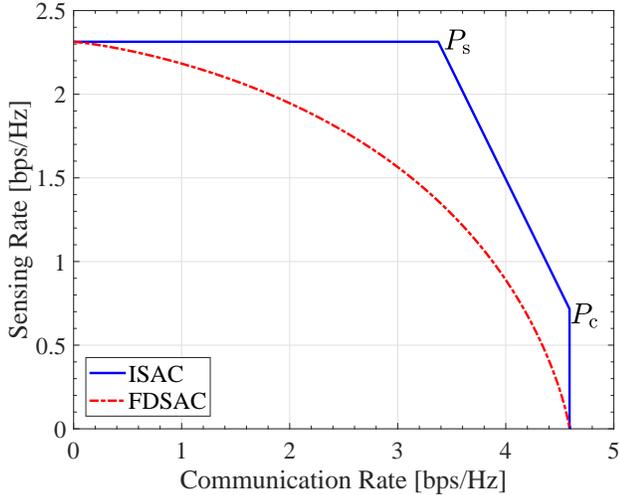}
\caption{Rate region achieved in uplink ISAC and FDSAC. The system parameters can be found in \cite{Ouyang2022_2}.}
\label{Uplink_Region}
\vspace{-10pt}
\end{figure}

The above argument suggests that the performance of uplink ISAC is influenced by the SIC order. To better illustrate this, in {\figurename} {\ref{Uplink_Region}}, we plot the rate region achieved by uplink ISAC. In obtaining {\figurename} {\ref{Uplink_Region}}, a minimum MSE (MMSE)-SIC decoder is used to detect the information bits sent by the uplink CUs, which is sum-rate capacity-achieving \cite{Ouyang2022_2}. Let us now focus on the points $P_{\rm{s}}$ and $P_{\rm{c}}$ in {\figurename} {\ref{Uplink_Region}}. Particularly, to achieve point $P_{\rm{s}}$, the BS should exploit the S-C SIC scheme, whereas to achieve point $P_{\rm{c}}$, the BS must exploit the C-C SIC scheme. It is worth noting that $P_{\rm{s}}$ and $P_{\rm{c}}$ achieve the largest SR and CR, respectively, which reflects the influence of the SIC order. The line segment connecting $P_{\rm{s}}$ and $P_{\rm{c}}$ is achieved by using the celebrated time-sharing strategy, namely applying the strategy corresponding to $P_{\rm{s}}$ with probability $p$, while applying the strategy corresponding to $P_{\rm{c}}$ with probability $1-p$. This line segment, essentially, represents a communication-sensing performance trade-off. In a nutshell, the BS can use SIC to achieve any point in the ISAC's rate region. As seen previously, SIC is at the heart of uplink ISAC, which achieves the best known S-C rate region in the uplink setting. For comparison, the rate region achieved by FDSAC is also presented in {\figurename} {\ref{Uplink_Region}}. As expected, it can be seen from {\figurename} {\ref{Uplink_Region}} that ISAC can achieve a broader rate region than FDSAC.

\subsection{High-SNR Slopes}

\begin{figure}[!t]
\centering
\setlength{\abovecaptionskip}{0pt}
\includegraphics[width=0.45\textwidth]{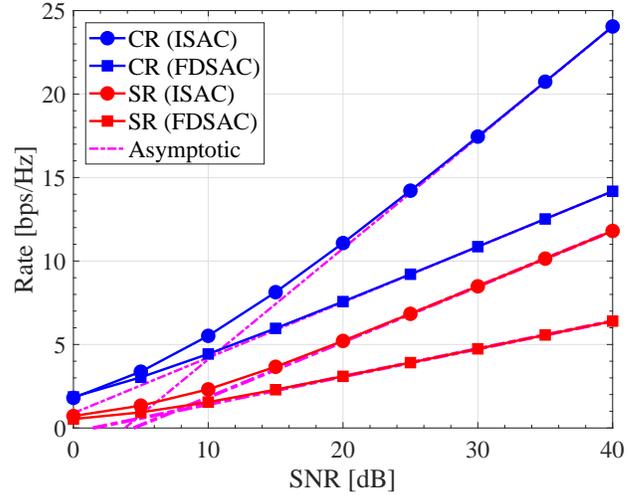}
\caption{Communication and sensing rates for uplink ISAC and FDSAC. The system parameters can be found in \cite{Ouyang2022_2}.}
\label{Uplink_Rate}
\vspace{-10pt}
\end{figure}

Having characterized the S-C rate region, we now move on to the high-SNR slopes. Thanks to the SIC technique, the uplink ISAC makes full use of all the spectrum resources in both communications and sensing. In comparison with FDSAC, the pre-log term in the CR/SR expression of ISAC will not be influenced by a spectrum allocation factor less than one. The uplink ISAC is therefore foreseen to enjoy larger high-SNR slopes than FDSAC. To show this benefit, in {\figurename} {\ref{Uplink_Rate}}, we compare FDSAC with the S-C SIC-based ISAC by presenting the corresponding CR and SR. By observing slopes of the asymptotic CR/SR curves, we find that ISAC is capable of achieving larger high-SNR slopes than FDSAC. Furthermore, as {\figurename} {\ref{Uplink_Rate}} shows, in the low-SNR regime, ISAC achieves virtually the same CR and SR as FDSAC. By contrast, in low and moderate SNR regions, ISAC supports a higher CR as well as SR than FDSAC.

Putting the above results together, we conclude that ISAC can outperform FDSAC in terms of both the S-C rate region and high-SNR slopes. That is to say, under the same resources, ISAC can support the transmission of more data information as well as the extraction of more environmental information than FDSAC. This superiority, essentially, originates from ISAC's integrated exploitation of spectrum and power resources.

\section{Conclusion and Promising Research Directions}
In this article, the performance of ISAC systems has been investigated from a MI perspective. We started by explaining the concept as well as the benefits of the MI-based framework. Then, two typical MI-related performance metrics, namely the S-C rate region and high-SNR slope, were highlighted. By using these two metrics as guidance, the performances of downlink and uplink ISAC were studied, followed by exploring the performance gap between ISAC and FDSAC. It is hoped that this article will help establish a unified as well as analytically tractable framework for ISAC and pave the way for designing dual-functional sensing-communication networks. There are still numerous open research problems in this area, some of which are listed below.
\begin{itemize}
  \item System-Level Performance Analysis: In addition to the S-C rate region and high-SNR slope, there are many other MI-related system-level performance metrics, such as the outage probability, diversity order, energy efficiency, transmission latency, and rate distortion function. Leveraging these metrics to evaluate the performance gap between ISAC and FDSAC is a promising research direction, which can unveil important system design insights.
  \item Network-Level Performance Analysis: In practice, ISAC may need to be deployed in a multi-cell environment. As the density for wireless networks increases, inter-cell interference becomes a major obstacle to achieving the benefits of ISAC. As such, analyzing the rate region and high-SNR slope from a network level and finding corresponding interference management approaches are of great significance. These problems require further exploration in future research.
  \item Pareto Boundary Characterization: Beamforming design has a significant impact on the performance of multiple-antenna-assisted ISAC. Since different qualities of communication and sensing services should be satisfied simultaneously, the beamforming design problem in ISAC is generally challenging. As such, finding low-complexity methods to approach the Pareto boundary of the S-C rate region makes important sense. Such Pareto-optimal designs still constitute an open area.
\end{itemize}

\begin{IEEEbiographynophoto}{Chongjun Ouyang} is currently pursuing the Ph.D. degree in Beijing University of Posts and Telecommunications. His research interests include multi-antenna technologies with an emphasis on ISAC.
\end{IEEEbiographynophoto}

\begin{IEEEbiographynophoto}{Yuanwei Liu} [Senior Member, IEEE] is a senior lecturer (associate professor) at Queen Mary University of London. His research interests include 5G/6G networks, NOMA, RIS, UAV communications, and machine learning. He is a Web of Science Highly Cited Researcher 2021. He is currently a Senior Editor of IEEE Communications Letters, and an Editor of IEEE Transactions on Wireless Communications and IEEE Transactions on Communications. He received the IEEE ComSoc Outstanding Young Researcher Award for EMEA in 2020. He received the 2020 IEEE Signal Processing and Computing for Communications Technical Early Achievement Award and the IEEE Communication Theory Technical Committee 2021 Early Achievement Award. He received the IEEE ComSoc Young Professional Outstanding Nominee Award in 2021.
\end{IEEEbiographynophoto}

\begin{IEEEbiographynophoto}{Hongwen Yang} [Member, IEEE] is a full professor with Beijing University of Posts and Telecommunications. His research mainly focuses on wireless physical layer, including modulation and coding, MIMO, and OFDM.
\end{IEEEbiographynophoto}

\begin{IEEEbiographynophoto}{Naofal Al-Dhahir} [Fellow, IEEE] is Erik Jonsson Distinguished Professor \& ECE Dept. Associate Head at UT-Dallas. He earned his PhD degree from Stanford University and was a principal member of technical staff at GE Research Center and AT\&T Shannon Laboratory from 1994 to 2003. He is co-inventor of 43 issued patents, co-author of about 500 papers and co-recipient of 5 IEEE best paper awards. He is an IEEE Fellow, received 2019 IEEE SPCC technical recognition award and 2021 Qualcomm faculty award. He served as Editor-in-Chief of IEEE Transactions on Communications from Jan. 2016 to Dec. 2019. He is a Fellow of the National Academy of Inventors.
\end{IEEEbiographynophoto}


\begin{thebibliography}{00}
\bibitem{Saad2019} W. Saad, M. Bennis, and M. Chen, ``A vision of 6G wireless systems: Applications, trends, technologies, and open research problems,'' \emph{IEEE Netw.}, vol. 34, no. 3, pp. 134--142, 2019.
\bibitem{Liu2022} F. Liu \emph{et al.}, ``Integrated sensing and communications: Towards dual-functional wireless networks for 6G and beyond,'' \emph{IEEE J. Sel. Areas Commun.}, vol. 40, no. 6, pp. 1728--1767, Jun. 2022.
\bibitem{Zhang2022} J. A. Zhang \emph{et al.}, ``Enabling joint communication and radar sensing in mobile networks--A survey,'' \emph{IEEE Commun. Surv. Tut.}, vol. 24, no. 1, pp. 306--345, 1st Quart. 2022.
\bibitem{Zhang2021} J. A. Zhang \emph{et al.}, ``An overview of signal processing techniques for joint communication and radar sensing,'' \emph{IEEE J. Sel. Topics Signal Process.}, vol. 15, no. 6, pp. 1295--1315, Nov. 2021.
\bibitem{Liu2021} A. Liu \emph{et al.}, ``A survey on fundamental limits of integrated sensing and communication,'' \emph{IEEE Commun. Surveys Tuts.}, vol. 24, no. 2, pp. 994--1034, 2nd Quart., 2022.
\bibitem{Liu2022_2} F. Liu \emph{et al.}, ``Cram\'{e}r-Rao bound optimization for joint radar-communication beamforming,'' \emph{IEEE Trans. Signal Process.}, vol. 70, pp. 240--253, 2022.
\bibitem{Yuan2021_1} W. Yuan \emph{et al.}, ''Bayesian predictive beamforming for vehicular networks: A low-overhead joint radar-communication approach'', \emph{IEEE Trans. Wireless Commun.}, vol. 20, no. 3, pp. 1442--1456, Mar. 2021.
\bibitem{Yuan2022_2} C. Liu \emph{et al.}, ``Learning-based predictive beamforming for integrated sensing and communication in vehicular networks,'' \emph{IEEE J. Sel. Areas Commun.}, vol. 40, no. 8, pp. 1978--1992, Aug. 2022.
\bibitem{Tse2005} D. Tse and P. Viswanath, \emph{Fundamentals of Wireless Communication}. Cambridge, U.K.: Cambridge Univ. Press, 2005.
\bibitem{Xing2020} C. Xing \emph{et al.}, ``New viewpoint and algorithms for water-filling solutions in wireless communications,'' \emph{IEEE Trans. Signal Process.}, vol. 68, pp. 1618--1634, 2020.
\bibitem{Tang2019} B. Tang and J. Li, ``Spectrally constrained MIMO radar waveform design based on mutual information,'' \emph{IEEE Trans. Signal Process.}, vol. 67, no. 3, pp. 821--834, Feb. 2019.
\bibitem{Ouyang2022_2} C. Ouyang, Y. Liu, and H. Yang, ``Performance of downlink and uplink integrated sensing and communications (ISAC) systems,'' \emph{IEEE Wireless Commun. Lett.}, early access, 2022.
\bibitem{Ouyang2022_1} C. Ouyang, Y. Liu, and H. Yang, ``NOMA-ISAC: Performance analysis and rate region characterization,'' Oneline: \url{https://arxiv.org/abs/2205.13756}, Accessed on 27 May 2022.
\bibitem{Liu2017} Y. Liu \emph{et al.}, ``Non-orthogonal multiple access for 5G and beyond,'' \emph{Proc. IEEE}, vol. 105, no. 12, pp. 2347--2381, Dec. 2017.
\bibitem{Ouyang2022_3} C. Ouyang, Y. Liu, and H. Yang, ``On the performance of uplink ISAC systems,'' \emph{IEEE Commun. Lett.}, early access, 2022.
\end{thebibliography}
 \end{document}